\documentclass[twocolumn,english,aps,pra,superscriptaddress,showpacs,tightenlines]{revtex4-1}
\usepackage{amsmath}
\usepackage{graphicx}
\usepackage{amssymb}
\usepackage{CJK}
\usepackage{color}

\begin{document}

\begin{CJK*}{GBK}{song}

\title{interference trapping of populations in a semi-infinite coupled-resonator waveguide}
\author{Qinglian \surname{Fu} }
\affiliation{Key Laboratory of Low-Dimension Quantum Structures and Quantum Control of Ministry of Education, Key Laboratory for Matter Microstructure and Function of Hunan Province, Synergetic Innovation Center for Quantum Effects and Applications, Xiangjiang-Laboratory and Department of Physics, Hunan Normal University, Changsha 410081, China}
\author{Jing \surname{Li} }
\affiliation{Key Laboratory of Low-Dimension Quantum Structures and Quantum Control of Ministry of Education, Key Laboratory for Matter Microstructure and Function of Hunan Province, Synergetic Innovation Center for Quantum Effects and Applications, Xiangjiang-Laboratory and Department of Physics, Hunan Normal University, Changsha 410081, China}
\affiliation{Institute of Interdisciplinary Studies, Hunan Normal University, Changsha, 410081, China}
\author{Jing \surname{Lu}}
\thanks{Corresponding author}
\email{lujing@hunnu.edu.cn}
\affiliation{Key Laboratory of Low-Dimension Quantum Structures and Quantum Control of Ministry of Education, Key Laboratory for Matter Microstructure and Function of Hunan Province, Synergetic Innovation Center for Quantum Effects and Applications, Xiangjiang-Laboratory and Department of Physics, Hunan Normal University, Changsha 410081, China}
\affiliation{Institute of Interdisciplinary Studies, Hunan Normal University, Changsha, 410081, China}
\author{Zhihui \surname{Peng}}
\affiliation{Key Laboratory of Low-Dimension Quantum Structures and Quantum Control of Ministry of Education, Key Laboratory for Matter Microstructure and Function of Hunan Province, Synergetic Innovation Center for Quantum Effects and Applications, Xiangjiang-Laboratory and Department of Physics, Hunan Normal University, Changsha 410081, China}
\affiliation{Institute of Interdisciplinary Studies, Hunan Normal University, Changsha, 410081, China}
\author{Lan \surname{Zhou}}
\affiliation{Key Laboratory of Low-Dimension Quantum Structures and Quantum Control of Ministry of Education, Key Laboratory for Matter Microstructure and Function of Hunan Province, Synergetic Innovation Center for Quantum Effects and Applications, Xiangjiang-Laboratory and Department of Physics, Hunan Normal University, Changsha 410081, China}
\affiliation{Institute of Interdisciplinary Studies, Hunan Normal University, Changsha, 410081, China}

\begin{abstract}
We study the energy structure and dynamics of a two-level emitter (2LE) locally coupled to
a semi-infinite one-dimensional (1D) coupled-resonator array (CRA). The energy spectrum in the
single-excitation subspace features a continuous band with scattering states, discrete levels
with bound states, and a quantum phase transition characterized by the change of the number of
bound states. The number of bound states is revealed by the behavior of the excited-state
population at long times with quantum beat, residual oscillation, a constant with either
non-zero or zero.
\end{abstract}
\pacs{}
\maketitle

\end{CJK*}\narrowtext

\section{Introduction}

Waveguide quantum electrodynamics (WQED) where quantum emitters (QEs)
interact with propagating photons in a one-dimensional (1D) waveguide, is promising for quantum networks and quantum computation. The 1D confinement of light makes
it possible for single photons (SPs) to be efficiently absorbed by even a
single emitter, the quantum interference between the incident wave and the
emitted ones gives rise to many potential applications, such as SP switches~%
\cite{ShenPRL95,LanPRL101}, SP routers~\cite%
{HoiPRL107,lanPRL111,LuPRA89,routPRA94,WeiPRA89,AhumPRA99,routPRR02}, SP
memory devices~\cite%
{LanPRA78,gongPRA78,DongPRA79,guoprr2,KianPRA107,PRA111ya,PRA111ge}, and so
on. A paradigmatic system in WQED is the 1D coupled-resonator array (CRA),
which typically is an arrangement of low-loss resonators with
nearest-neighbor coupling and allowing photon hopping between adjacent
resonators. Resonators in such array are handled as individual sites, so a
1D CRA is usually described by the tight-bind model\cite%
{lanPRL111,LuPRA89,LanPRA85,PRA95xu,PRL127You,PRR4Jin,RupakPRA108,PRA109Da}.
A QE coupled to a CRA has a continuous band and out-of band discrete levels%
\cite%
{LanPRA80,SEPRA89,SEPRA96,PRA100LQ,wangPRA101,NoriPRL126,PRX12,LiNJP26,luOL49}%
, where the former associate unbound stationary states and the latter
correspond to bound states outside of the continuum (BOCs).

The CRA usually is regarded as endless, however, a CRA with a single-end is
also possible. Attentions are paid on a QE coupled to such semi-infinite CRA
since the some amount of radiation emitted by the QE and back reflected by
the end to the QE~\cite{LuOE23,PRA109Jun}. In this paper, we consider a
point-like two-level emitter (2LE) interacting with one of single resonator in a semi-infinite CRA. By analyzing energy spectrum in
one single-excitation subspace, besides the quantum phase transition
characterized by the number of BOCs~\cite{luOL49,LeiPRA100}, a bound
stationary state that arises within a continuum of unbound states is found.
This bound state in the continuum (BIC) is a dressed state consisting of a 2LE coupled to a single photon, which is strictly confined within the region between the 2LE and the end of the structure. Remarkably, this state has exactly the same energy as the bare 2LE. Then, we studied the emission process of a 2LE in this
semi-infinite CRA. Owing to the number of the bound states, the
excited-state population at long times displays a behavior with quantum
beat, residual oscillatory, or non-zero constant.

\section{\label{Sec:2}Setup and Hamiltonian}

\begin{figure}[tbph]
\includegraphics[width=8 cm,clip]{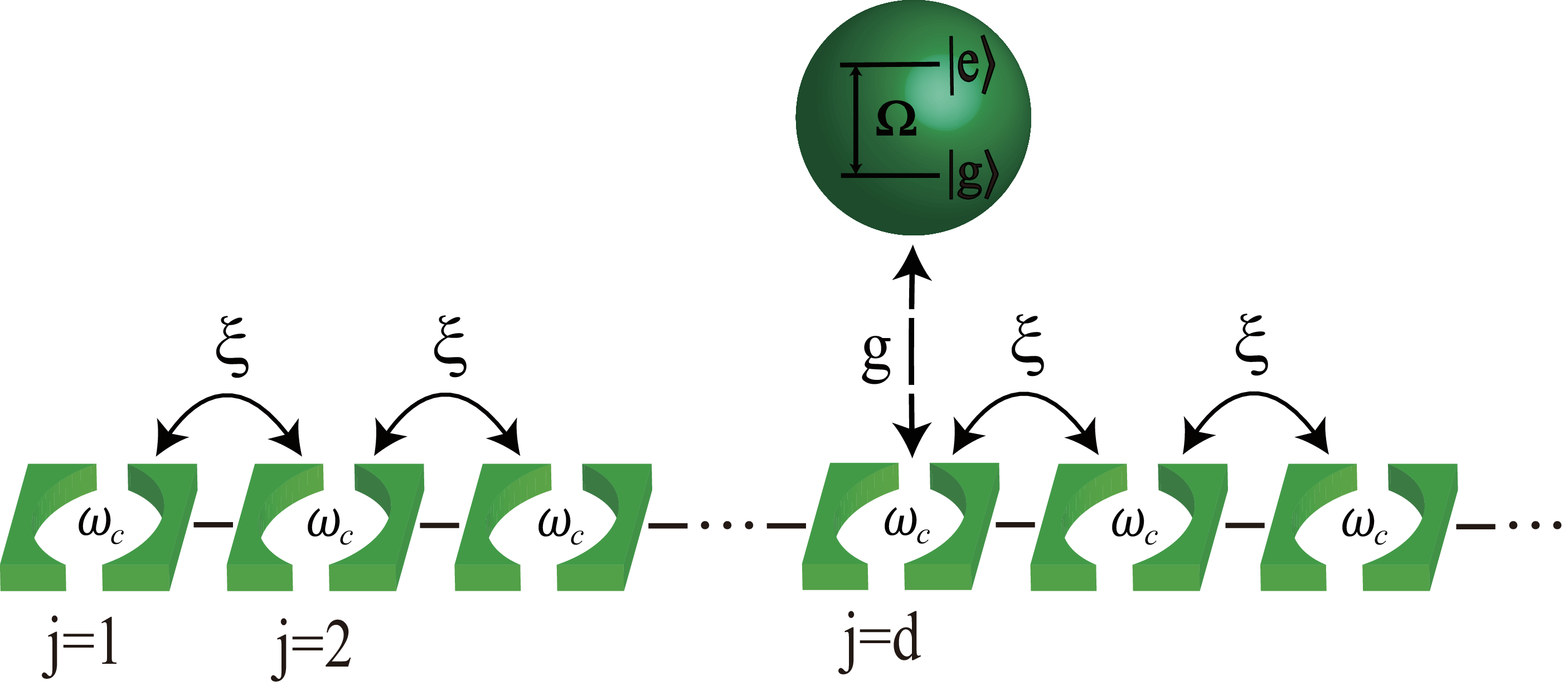}
\caption{Sketch of the system: a two-level quantum emitter and a
semi-infinite 1D waveguide made of coupled single-mode resonator with
nearest-neighbor coupling $\protect\xi $, where the $d$th resonator is
coupled to an initially excited two-level atom.}
\label{fig1}
\end{figure}
A 2LE with energy space $\Omega $ between the excited state $\left\vert
e\right\rangle $ and the ground state $\left\vert g\right\rangle $ weakly
coupled to a 1D semi-infinite CRA as shown in Fig.~\ref{fig1}. The full
2LE-CRA Hamiltonian is $\hat{H} = \hat{H}_{a} + \hat{H}_{w} + \hat{H}_{i}$, where the $\hat{H}_{a}=\Omega \left\vert e\right\rangle \left\langle e\right\vert $ is the free Hamiltonian of the 2LE. The waveguide is modeled as a tight-binding array of single-mode identical coupled resonators with Hamiltonian (hereafter $\hbar=1$)%
\begin{equation}  \label{Eq1-01}
\hat{H}_{w}=\omega _{c}\sum_{j}\hat{a}_{j}^{\dagger }\hat{a}_{j}-\xi
\sum_{j}\left( \hat{a}_{j+1}^{\dagger }\hat{a}_{j}+\hat{a}_{j}^{\dagger }%
\hat{a}_{j+1}\right),
\end{equation}%
where $\omega _{c}$ is the resonance frequency of each cavity, $\hat{a}_{j}$
($\hat{a}_{j}^{\dagger }$) are real space bosonic annihilation (creation)
operators at $j$-th resonator, and $\xi $ is the coupling strength between
neighboring resonators. The Fourier transformation $\hat{a}_{k}=\sqrt{2/N}%
\sum_{j}\hat{a}_{j}\sin \left( kj\right) $ diagonalize the free waveguide Hamiltonian into $\hat{H}_{w}=\sum_{k}\omega _{k}%
\hat{a}_{k}^{\dagger }\hat{a}_{k}$ with $\omega _{k}=\omega _{c}-2\xi \cos k$
(The lattice constant is set to $1$). The interaction Hamiltonian within the
rotating-wave approximation $\hat{H}_{i}=g\hat{a}_{d}^{\dagger }\hat{\sigma}_{-}+h.c.$ describes the local coupling of a 2LE to the resonator $j=d$. In momentum space,
the interaction Hamiltonian reads
\begin{equation}
\label{Eq1-02}
\hat{H}_{i}=\sum_{k}\left( g_{k}\hat{a}_{k}^{\dagger }\hat{\sigma}_{-}+h.c.\right)
\end{equation}%
and $g_{k}=g\sqrt{2/N}\sin \left( kd\right) $.

\section{Single-Excitation Spectrum}

\begin{figure}[tbph]
\includegraphics[width=8 cm,clip]{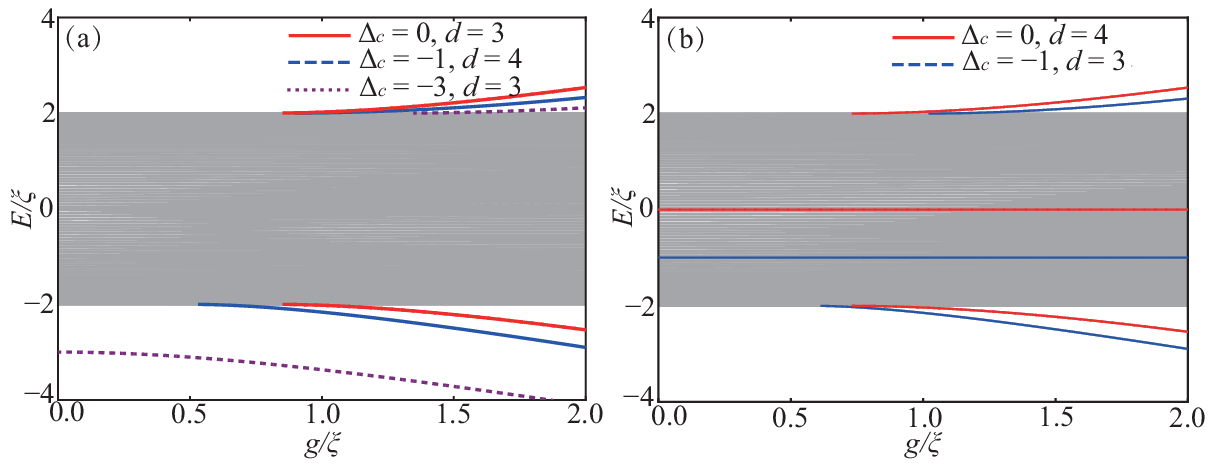}
\caption{The single-excitation spectrum as a function of the coupling
strength. The parameters are (a) $\Delta _{c}=0, d=3$ (red solid lines), $\Delta _{c}=-1, d=4$ (blue dashed lines) and $\Delta _{c}=-3, d=3$ (purple dotted lines); (b) $\Delta _{c}=0, d=4$ (red solid lines) and $\Delta _{c}=-1, d=3$ (blue dashed
lines), respectively. The parameters are in units of the hop strength $\protect\xi $, so do the following figures. }
\label{fig2}
\end{figure}
The 2LE-waveguide Hamiltonian $H$ entails conservation of the total number
of excitations. In the single-excitation subspace, the state vector of
the system at an arbitrary time
\begin{equation}
\left\vert \psi \left( t\right) \right\rangle =\sum_{k}A_{k}\left( t\right)
\hat{a}_{k}^{\dagger }\left\vert g0\right\rangle +u\left( t\right)
\left\vert e0\right\rangle   \label{Eq2-01}
\end{equation}%
is superpositions of an 2LE's excitation $\left\vert e0\right\rangle $ and
single-photon states $\hat{a}_{k}^{\dagger }\left\vert g0\right\rangle $,
where state $\left\vert e0\right\rangle $ describes the 2LE in its excited
state $\left\vert e\right\rangle $ and the CRA in vacuum, and the state
vector $\hat{a}_{k}^{\dagger }\left\vert g0\right\rangle $ describes the 2LE
in the ground state $\left\vert g\right\rangle $ and a single photon in a
mode $k$. The time-dependent Schroedinger equation gives the dynamics of the
total system as
\begin{subequations}
\label{Eq2-02}
\begin{eqnarray}
\partial _{t}u\left( t\right)  &=&-i\Omega u\left( t\right) -%
i\sum_{k}g_{k}^{\ast }A_{k}\left( t\right) , \\
\partial _{t}A_{k}\left( t\right)  &=&-i\omega _{k}A_{k}\left(
t\right) -ig_{k}u\left( t\right).
\end{eqnarray}%
By applying the Fourier transform, the eigenenergy satisfies the
transcendental equation
\end{subequations}
\begin{equation}
E=\Omega +\frac{\left\vert g\right\vert ^{2}}{2\pi }\int_{-\pi }^{\pi }dk%
\frac{1-\cos \left( 2kd\right) }{E-\omega _{k}}. \label{Eq2-03}
\end{equation}%
It has two types of solutions: a continuous band and discrete levels. The
continuous band with $E_{k}=\omega _{k}\in \left[ \omega _{c}-2\xi ,\omega
_{c}+2\xi \right] $ associates scattering states
\begin{eqnarray}
\label{Eq2-04}
\left\vert E_{k}\right\rangle  &=&C_{k}\left\vert e0\right\rangle
+\sum_{j}A_{k}\sin \left( kj\right) \theta \left( d-j\right) \hat{a}_{j}^{\dagger }\left\vert g0\right\rangle   \\
&&+\sum_{j}\left[ e^{-ik\left( j-d\right) }+r_{k}e^{ik\left( j-d\right) }%
\right] \theta \left( j-d\right) \hat{a}_{j}^{\dagger }\left\vert g0\right\rangle,\notag
\end{eqnarray}%
spatially extending over the whole waveguide and
\begin{subequations}
\label{Eq2-04}
\begin{eqnarray}
r_{k} &=&-\frac{\left\vert g\right\vert ^{2}\sin \left( kd\right) +\xi
\left( E_{k}-\Omega \right) e^{ikd}\sin k}{\left\vert g\right\vert ^{2}\sin
\left( kd\right) +\xi \left( E_{k}-\Omega \right) e^{-ikd}\sin k} ,\\
C_{k} &=&\frac{-2ig^{\ast }\xi \sin k\sin \left( kd\right) }{\left\vert
g\right\vert ^{2}\sin \left( kd\right) +\xi \left( E_{k}-\Omega \right) e^{-%
ikd}\sin k}, \\
A_{k} &=&\frac{-2i\xi \left( E_{k}-\Omega \right) \sin k}{%
\left\vert g\right\vert ^{2}\sin \left( kd\right) +\xi \left( E_{k}-\Omega
\right) e^{-ikd}\sin k}.
\end{eqnarray}%
Discrete levels associate 2LE-photon bound states with a photon localized in
a finite regime. In Fig.~\ref{fig2}, we plot the energy spectrum as a
function of the coupling strength $g$ for a given detuning $\Delta
_{c}=\Omega -\omega _{c}$ by numerically diagonalizing the Hamiltonian with $N=102$. Besides the continuous band shown by the shaded region, we observe
two out-of-band discrete levels\cite%
{LanPRA80,SEPRA89,SEPRA96,PRA100LQ,wangPRA101,NoriPRL126,PRX12,LiNJP26,luOL49}
denoted as $E_{u}=\omega _{c}+2\xi \sinh \kappa _{u}$ and $E_{l}=\omega
_{c}-2\xi \sinh \kappa _{l}$ with corresponding bound states outside of the
continuum (BOCs)
\end{subequations}
\begin{subequations}
\label{Eq2-05}
\begin{eqnarray}
\left\vert E_{u}\right\rangle  &=&C_{u}\left\vert e0\right\rangle
+\sum_{j}A_{u}\sinh \left( \kappa _{u}j\right) \theta \left( d-j\right)
\left( -1\right) ^{j}\hat{a}_{j}^{\dagger }\left\vert g0\right\rangle  \\
&&+\sum_{j}A_{u}\sinh \left( \kappa _{u}d\right) e^{-\kappa _{u}\left(
j-d\right) }\theta \left( j-d\right) \left( -1\right) ^{j}\hat{a}%
_{j}^{\dagger }\left\vert g0\right\rangle, \notag \\
\left\vert E_{l}\right\rangle  &=&C_{l}\left\vert e0\right\rangle
+\sum_{j}A_{l}\sinh \left( \kappa _{l}j\right) \theta \left( d-j\right) \hat{%
a}_{j}^{\dagger }\left\vert g0\right\rangle  \\
&&+\sum_{j}A_{l}\sinh \left( \kappa _{l}d\right) e^{-\kappa _{l}\left(
j-d\right) }\theta \left( j-d\right) \hat{a}_{j}^{\dagger }\left\vert
g0\right\rangle \notag.
\end{eqnarray}
\end{subequations}
Here,
\begin{subequations}
\label{Eq2-06}
\begin{eqnarray}
C_{u} &=&\left( -1\right) ^{d}\frac{g^{\ast }\left( e^{\kappa
_{u}d}-e^{-\kappa _{u}d}\right) }{2\left( E-\Omega \right) }A_{u} ,\\
C_{l} &=&\frac{g^{\ast }\left( e^{\kappa _{l}d}-e^{-\kappa _{l}d}\right) }{%
2\left( E-\Omega \right) }A_{l} ,\\
A_{\alpha }^{-1} &=&\sqrt{\frac{\cosh \kappa _{\alpha }\left( e^{2\kappa
_{\alpha }d}-1\right) }{4\sinh \kappa _{\alpha }}+\left\vert g\right\vert
^{2}\frac{\sinh ^{2}\left( \kappa _{\alpha }d\right) }{\left( E_{\alpha
}-\Omega \right) ^{2}}-\frac{d}{2}}.
\end{eqnarray}
\end{subequations}
Energies $E_{\alpha}, \alpha=u,l$ move away from the band as $g$ increases.
When $\Delta _{c}=0$, the energies $E_{u}$ and $E_{l}$ are symmetrically
located around the band, and $C_{u}=(-1)^{d}C_{l}$.  As $\Omega $ increases
(decreases) from $\omega _{c}$, that energy $E_{u}$ ($E_{l}$) moves away from
the band more rapidly than energy $E_{l}$ ($E_{u}$) as $g$ increases. Similar
results are obtained in Ref.\cite{SEPRA96,luOL49}. When $\Omega <\omega _{c}-2\xi $
($\Omega >\omega _{c}+2\xi $), $E_{l}$ ($E_{u}$) always exists no matter what
values $g$ take, otherwise, the energies of BOCs are dependent on $g$, so the
number of BOCs is controllable. To understand the appearance and disappearance
of two BOCs, we integrate the second term on the right-hand of Eq.(\ref{Eq2-03}),
and obtain
\begin{subequations}
\label{Eq2-07}
\begin{eqnarray}
&&\left( -E_{u}+\Omega \right) \sqrt{\left( \frac{E_{u}-\omega _{c}}{2\xi }%
\right) ^{2}-1}  \notag \\
&=&-\frac{\left\vert g\right\vert ^{2}}{2\xi }+\frac{\left\vert g\right\vert
^{2}}{2\xi }\left[ -\frac{E_{u}-\omega _{c}}{2\xi }+\sqrt{\left( \frac{%
E_{u}-\omega _{c}}{2\xi }\right) ^{2}-1}\right] ^{2d}, \\
&&\left( -E_{l}+\Omega \right) \sqrt{\left( \frac{E_{l}-\omega _{c}}{2\xi }%
\right) ^{2}-1}  \notag \\
&=&\frac{\left\vert g\right\vert ^{2}}{2\xi }-\frac{\left\vert g\right\vert
^{2}}{2\xi }\left[ -\frac{E_{l}-\omega _{c}}{2\xi }-\sqrt{\left( \frac{%
E_{l}-\omega _{c}}{2\xi }\right) ^{2}-1}\right] ^{2d}
\end{eqnarray}%
for a BOC in the upper and lower bands, respectively. We found that the
upper (lower) BOC occurs at the condition $g>g_{u}=\sqrt{\frac{2\xi ^{2}-\xi
\Delta_{c} }{d}}$ ($g>g_{l}=\sqrt{\frac{2\xi ^{2}+\xi
\Delta_{c} }{d}}$) for $\Omega <\omega _{c}+2\xi $ ($%
\Omega >\omega _{c}-2\xi $). We further observe a discrete level $%
E_{I}=\omega _{K}$ with wavenumber $K$ for $\Omega $ inside the band as long
as $g\neq 0$ (see Fig.\ref{fig2}b). The level $E_{I}$ associates a bound
state in the continuum (BIC)
\end{subequations}
\begin{equation}
\left\vert E_{I}\right\rangle =\sum_{j}A_{I}\sin \left( Kj\right) \theta
\left( d-j\right) \hat{a}_{j}^{\dagger }\left\vert g0\right\rangle
+C_{I}\left\vert e0\right\rangle ,   \label{Eq2-08}
\end{equation}%
and the emergence of the BIC is determined by the condition $\sin \left(
Kd\right) =0$ and $\omega _{K}=\Omega $. Here,
\begin{subequations}
\label{Eq2-09}
\begin{eqnarray}
C_{I} &=&\sqrt{\frac{2\xi ^{2}\sin ^{2}K}{2\xi ^{2}\sin ^{2}K+d{\left\vert
g\right\vert ^{2}}\cos ^{2}\left( Kd\right) }}, \\
A_{I} &=&\frac{-g\cos \left( Kd\right) }{\sqrt{\xi ^{2}\sin ^{2}K+d{%
\left\vert g\right\vert ^{2}}\cos ^{2}\left( Kd\right) /2}}.
\end{eqnarray}
\end{subequations}

\section{Atomic Emission Dynamics}

\begin{figure}[tbph]
\includegraphics[width=8 cm,clip]{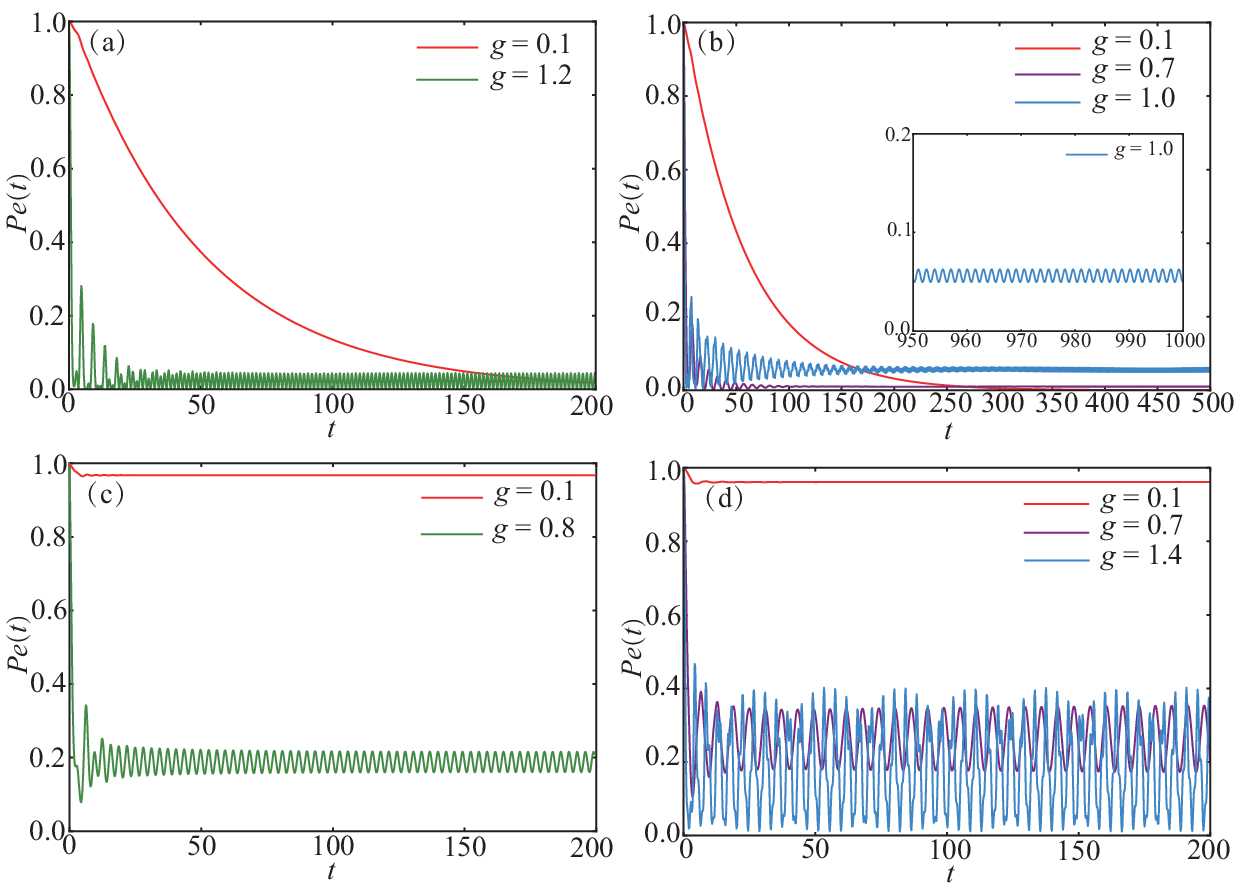}
\caption{Time evolution of atomic excitation probability $P_{e}(t)$. We set (a)
$\Delta_c=0, d=3$, (b) $\Delta_c=-1, d=4$, (c) $\Delta_c=0, d=4$, (d) $%
\Delta_c=-1, d=3$. }
\label{fig3}
\end{figure}

Consider the dynamics of the system initially in state $\left\vert \psi
\left( 0\right) \right\rangle =\left\vert e0\right\rangle $, i.e.,
investigate the atomic emission into the vacuum field by an initially
excited 2LE. With bound and unbound stationary states all given in the last
section, the 2lE's probability amplitude reads
\begin{eqnarray}
\label{Eq2-10}
u\left( t\right) &=&\int \frac{dk}{2\pi }e^{-\mathrm{i}E_{k}t}\left\vert
C_{k}\right\vert ^{2}+e^{-\mathrm{i}E_{u}t}\left\vert C_{u}\right\vert ^{2}
\notag \\
&&+e^{-\mathrm{i}E_{l}t}\left\vert C_{l}\right\vert ^{2}+e^{-\mathrm{i}%
E_{I}t}\left\vert C_{I}\right\vert ^{2},
\end{eqnarray}%
where $C_{k}$, $C_{u}$, $C_{l}$, and $C_{I}$ are the amplitudes for the 2LE
to be excited in the unbound state $\left\vert E_{k}\right\rangle $ and
bound states $\left\vert E_{u}\right\rangle $, $\left\vert
E_{l}\right\rangle $, and $\left\vert E_{I}\right\rangle $, respectively. To
find the contributions of the unbound and bound states, we plot the 2LE's
probability $P_{e}\left( t\right) =\left\vert u\left( t\right) \right\vert
^{2}$ in Fig.\ref{fig3} for different coupling strengths $g$ and detunings $%
\Delta _{c}$ with the transition frequency $\Omega \in \left[ \omega
_{c}-2\xi ,\omega _{c}+2\xi \right] $. The BIC is absent for Fig.\ref{fig3}%
(a,b), but the BIC is present for Fig.\ref{fig3}(c,d). When $g<\min \left(
g_{u},g_{l}\right) $, both BIC and BOCs are absent for the red lines in Fig.%
\ref{fig3}(a,b), so the initial excitation in the 2LE is entirely released
to the field and is distributed along the waveguide. However, considerable
populations reside in level $|e\rangle$ owing to the presence of the BIC for
red lines in Fig.\ref{fig3}(c,d). Since two BOCs have amplitudes $%
C_{u}=C_{l} $ and energies $\left\vert E_{u}-E_{I}\right\vert =\left\vert
E_{l}-E_{I}\right\vert $ when $g>\max\left( g_{u},g_{l}\right) $ and $\Delta
_{c}=0$, the probability $P_{e}\left( t\right) $ exhibits rapid oscillatory
decay first and then oscillates with fixed frequency $\left\vert
E_{u}-E_{I}\right\vert $ in the long term, see the green lines in Fig.\ref%
{fig3}(a,c), and $\min P_{e}\left( t\right) \neq 0$ in Fig.\ref{fig3}(c)
indicates the energy trapping by the BIC due to $C_{I}\neq 0$. When detuning
$\Delta _{c}<0$ and $\min \left( g_{u},g_{l}\right) <g<\max \left(
g_{u},g_{l}\right) $, only BOC $\left\vert E_{l}\right\rangle $ exists, so a
constant $P_{e}\left( t\right) $ can be observed after sufficiently long
time, see the purple line in Fig.\ref{fig3}(b). However, there are a BOC $%
\left\vert E_{l}\right\rangle $ and a BIC $\left\vert E_{I}\right\rangle $
for the purple line in Fig.\ref{fig3}(d). One observed $P_{e}\left( t\right)
$ oscillating with frequency $\left\vert E_{l}-E_{I}\right\vert $ and some
amount of population trapping within the 2LE after sufficiently long time.
As $g$ increases till $g>\max \left( g_{u},g_{l}\right) $, a stationary
oscillation at long enough times is found since the system only have two
BOCs for the blue line in Fig.\ref{fig3}(b). However, two BOCs and a BIC are
present for the blue line in Fig.\ref{fig3}(d). The amplitudes $C_{u}\neq
C_{l}$ and energies $\left\vert E_{u}-E_{I}\right\vert \neq \left\vert
E_{l}-E_{I}\right\vert $ due to $\Delta _{c}\neq 0$ produce a quantum beat
of $P_{e}\left( t\right) $ in the long-time limit. As the coupling strength
increases, the trapping probability grows in Figs.\ref{fig3}(a,b), but
decreases in Fig.\ref{fig3}(c,d).

\section{Conclusion}

In this paper, we consider a point-like 2LE electric-dipole coupled to a
semi-infinite 1D CRA. The energy spectrum in the single-excitation subspace
consists of the continuum band of scattering states and three possible
discrete levels of two BOCs and one BIC, however, the number of discrete
levels is changed when the coupling strength, the transition frequency of
the 2LE and the end-2LE distance vary, which reveals a quantum phase
transition. The time evolution of the 2LE's excitation are then studied
after all stationary states are presented. Since the energies of two BOCs are symmetrically located around the bound when the emitter's transition frequency is resonant with the resonator, the 2LE's population at long times shows two distinct behaviors. In the absence of the BIC, it either decays to zero or persistently oscillates with low probability. In the presence of the BIC, it either keeps large in amount or persistently oscillates with an intermediate value. These behaviors occur as long as the 2LE is close to the end.
When $\Omega\neq \omega_c$, one bound state traps the excitation in a nonzero constant, the co-existance of two bound state
leads to the behavior of the 2LE's population with stationary oscillation.
Furthermore, the appearance of both BOCs and the BIC is manifested in the form of
quantum beat of the 2LE's population in the long time limit.

\begin{acknowledgments}
This work was supported by NSFC Grants No. 11935006, No. 12075082, No. 12247105, No. 92365209, No. 12421005, the science and technology innovation Program of Hunan Province (Grant No. 2020RC4047), XJ-Lab Key Project (23XJ02001), and the Science $\And $ Technology Department of Hunan Provincial Program (2023ZJ1010).
\end{acknowledgments}

\end{document}